\documentclass[lettersize,journal]{IEEEtran}
\usepackage{amsmath,amsfonts}
\usepackage{algorithmic}
\usepackage{algorithm}
\usepackage{array}
\usepackage[caption=false,font=normalsize,labelfont=sf,textfont=sf]{subfig}
\usepackage{textcomp}
\usepackage{stfloats}
\usepackage{url}
\usepackage{verbatim}
\usepackage{graphicx}
\usepackage{cite}
\usepackage{color}
\newtheorem{assumption}{Assumption}
\begin{document}

\title{Emergence of fairness behavior driven by reputation-based voluntary participation in evolutionary dictator games}

\author{Yanling Zhang, Yin Li, Xiaojie Chen, Guangming Xie
\thanks{This research was supported by the National Natural Science Foundation of China (Grants Nos. 62033010, 62036002, 61603036, 61976048). (Corresponding author: Xiaojie Chen or Guangming Xie).}	
\thanks{Y. Zhang and Y. Li are with School of Intelligence Science and Technology, University of Science and Technology Beijing, Beijing 100083, China}
\thanks{Y. Zhang and Y. Li are with Key Laboratory of Intelligent Bionic Unmanned Systems, Ministry of Education, University of Science and Technology Beijing, Beijing 100083, China}
\thanks{X. Chen is with School of Mathematical Sciences, University of Electronic Science and Technology of China, Chengdu 611731, China (email: xiaojiechen@uestc.edu.cn)}
\thanks{G. Xie is with Biomimetic Design Lab, College of Engineering, Peking University, Beijing 100871, China (email: xiegming@pku.edu.cn)}
}

\markboth{IEEE Transactions on Computational Social Systems}%
{Shell \MakeLowercase{\textit{et al.}}: A Sample Article Using IEEEtran.cls for IEEE Journals}


\maketitle

\begin{abstract}
Recently, reputation-based indirect reciprocity has been widely applied to the study on fairness behavior. Previous works mainly investigate indirect reciprocity by considering compulsory participation. While in reality, individuals may choose voluntary participation according to the opponent's reputation. It is still unclear how such reputation-based voluntary participation influences the evolution of fairness. To address this question, we introduce indirect reciprocity with voluntary participation into the dictator game (DG). 
We respectively consider good dictators or recipients can voluntarily participate in games when the opponents are assessed as bad. We theoretically calculate the fairness level under all social norms of third-order information. Our findings reveal that several social norms induce the high fairness level in both scenarios. However, more social norms lead to a high fairness level for voluntary participation of recipients, compared with the one of good dictators. The results also hold when the probability of voluntary participation is not low. Our results demonstrate that recipients' voluntary participation is more effective in promoting the emergence of fairness behavior.
\end{abstract}


\section{Introduction}
\IEEEPARstart{F} {airness} behavior, as a prominent prosocial behavior, has become a focal point in addressing social dilemmas between interacting agents~\cite{Chen2018,Wang2010,Valerio2021,Szolnoki2012}. Despite its social desirability, fairness behavior often involves personal sacrifices to provide benefits for others. Then the preference for unfair behavior is more favorable among individuals. Hence the study of fairness behavior from an evolutionary perspective has been a paradoxical challenge~\cite{Weiland2012,Boesch2019,Ambrose2019}. These circumstances raise one crucial question: How does fairness evolve? Over the past three decades, researchers have extensively investigated fairness in the context of the DG to address this question~\cite{Szolnoki2012a,Jeffrey2015,Snellman2019}.

In the DG, a dictator determines how to split a sum of money, and the division has to be accepted by the recipient. Game theory predicts that rational dictators in the DG would offer nothing to recipients in order to maximize their own payoffs. However, a DG experiment~\cite{Henrich2010}, which was conducted across $15$ diverse populations, revealed dictators provide a mean offer of $37\%$ for recipients. Similar results were also found in an another DG experiment~\cite{Engel2011}. Furthermore, recent evidence~\cite{Cappelen2016} implied that fairness behavior is intuitively ingrained in most people, indicating that acting fairly is a fundamental human trait. One plausible explanation for this phenomenon is indirect reciprocity. The mechanism means that players can obtain a positive reputation by cooperating with others. The human capacity of establishing reputation systems may provide valuable insights into the nature of fairness behavior~\cite{Samu2020,Roberts2021}.

Evolutionary game theory provides an efficient mathematical framework for studying indirect reciprocal models~\cite{Perc2013,Wang2015,Perc_2017_pr,Su2022,Hu2021,Deonauth2021,Han2023}. The theory shows that the strategy with higher fitness has a greater likelihood of spreading among individuals. In recent years, indirect reciprocal models have been extensively applied to the donation game~\cite{Clark2020,Schmid2021,Xia2023}. Social norms, serving as the foundation of indirect reciprocity, establish standards for assessing individuals' reputation~\cite{Santos2021,Capraro2019,Fehr2018}. Image scoring, as a basic first-order social norm, was widely investigated~\cite{Nowak1998,Berger2016}. The two-valued image scoring categorizes donors offering help as good and those refusing help as bad. However, this social norm results in the `scoring dilemma'~\cite{Panchanathan2004}. Specifically, justified defection can degrade discriminators' reputation, and discriminators will be rejected by other discriminators. Here, unconditional cooperation can invade the discriminating strategy, and unconditional defection can invade unconditional cooperation. In this way, defection dominates in the population. To resolve the `scoring dilemma', social norms of the second-order information were introduced to promote the evolution of cooperation~\cite{Suzuki2006}. The second-order information includes previous reputation of recipients and actions of donors. Later, social norms of the third-order information were established to achieve the stable cooperative behavior~\cite{Murase2022}. The third-order information includes actions of donors and previous reputation of donors and recipients.

How indirect reciprocity affects the emergence of fairness has been studied in the DG~\cite{Li2022}, by considering several particular social norms. In this study, we make a comprehensive exploration about the DG under all social norms of third-order information. As the benchmark, we focus on the case where it is compulsory for both the dictator and the recipient to participate in the game. 
Then voluntary participation is introduced into our model. Voluntary participation refers to the players' decision to abstain from participating in a game. This usually leads to minimal or no benefits for players. While most of previous studies have typically assumed that all players must participate in the game, real-life scenarios often allow players to choose whether to participate or not. An experiment also revealed that under certain circumstances, players prefer to adopt a loner strategy to obtain potentially lower but guaranteed returns, rather than pursue higher but risky returns~\cite{Hauert2005}. This further confirms the validity of voluntary participation. Behavioral experiments~\cite{Semmann2003} and theoretical researches~\cite{Hauert2002,Sasaki2012,Jia2018,Hu2019} have also demonstrated that the option of voluntary participation promotes the emergence of cooperation in the public goods game. Moreover, the option of voluntary participation has also been found to support the cooperative behavior in the prisoner's dilemma game~\cite{Iwamura2020,Krellner2023}.

In this paper, we focus on two research questions: How does reputation-based voluntary participation influence the evolution of fairness? Which social norms promote the emergence of fairness? We thereby consider two scenarios with the option of voluntary participation. In scenario $1$, good dictators are endowed with the opportunity of abstaining from games when recipients' reputation is deemed bad. In scenario $2$, recipients are endowed with the option of quitting from games with bad dictators. We calculate the fairness level under $256$ social norms. Then we explain the high fairness level by analyzing the pairwise competition between any two strategies and the reputation system in each monomorphic population. The main results can be summarized as follows.
	\begin{itemize}
		\item{ For compulsory participation, reputation alone cannot promote the emergence of fairness. }
		\item{ For voluntary participation of good dictators, fairness behavior can massively emerge for ten social norms.}
		\item {For voluntary participation of recipients, the evolution of fairness can be favored for sixteen social norms.}
\end{itemize}

\section{System model and method}
\begin{figure}[h]
	\centering
	\includegraphics[width=\linewidth]{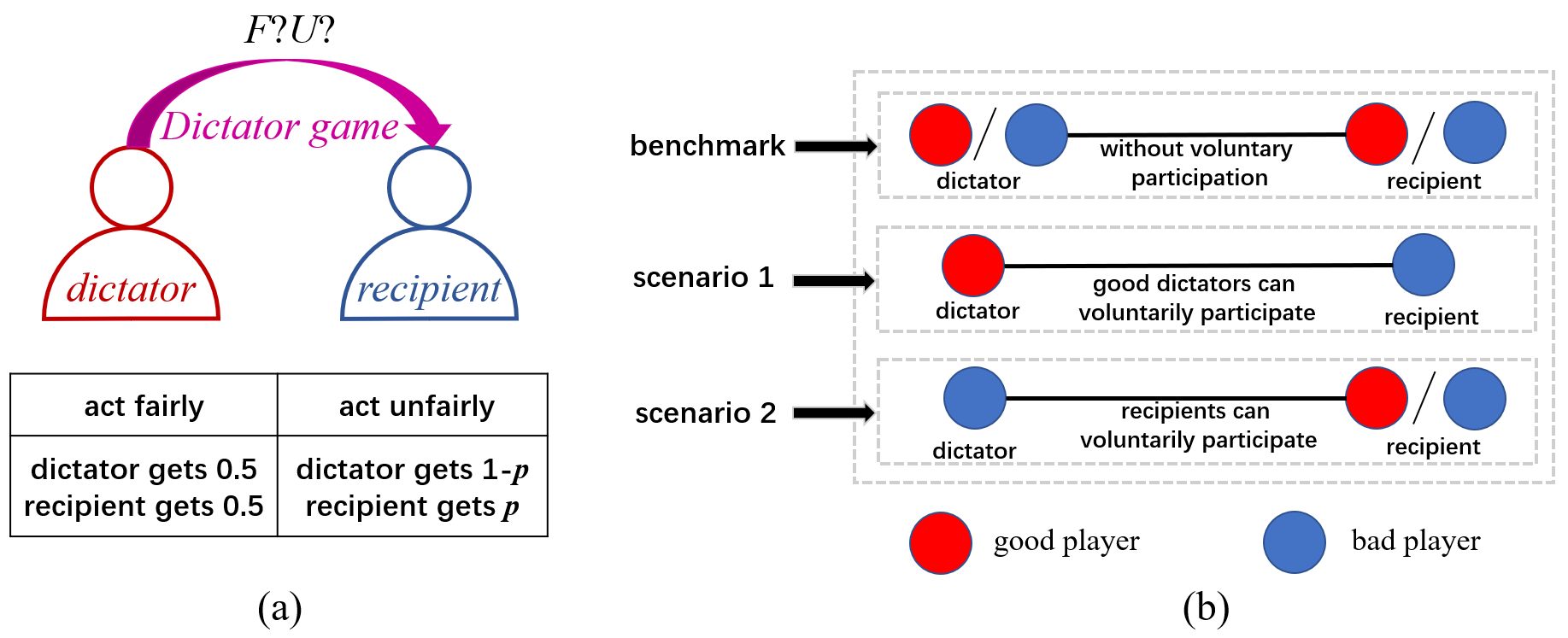}
	\caption{\label{f1} The schematic diagram of the model. (a) In the DG, the dictator acts fairly (F, equal distribution) or unfairly (U, offer $p\rightarrow 0$), and the recipient accepts the split unconditionally. (b) At the benchmark, neither the dictator nor the recipient has the option of voluntary participation. In the scenario $1$, good dictators abstain from games with a probability $p_{1}$ when the recipients have bad reputation. In the scenario $2$, recipients quit from games with a probability $p_{2}$ when the dictators have bad reputation.}
\end{figure}

We focus on a well-mixed population which is comprised of $Z$ players. In each round, two randomly chosen players participate in a simplified version of DG. As shown in Fig.~\ref{f1}(a), the dictator has to divide a predetermined sum with the recipient. The recipient is obliged to accept the division. The dictator has two strategies: the fair division ($F$) and the unfair division ($U$). The fair division denotes an equal distribution of the sum between the dictator and the recipient. Yet the unfair division indicates that the dictator offers an extremely low amount ($p \rightarrow 0$) to the recipient.

A player possesses a binary reputation value, either good ($G$) or bad ($B$). The dictator divides the money fairly or unfairly based on the reputation of the recipient. Then each player has the strategy represented by two letters $s_{G}s_{B}$. The first letter ($s_G$) indicates whether the dictator divides the money fairly ($s_G=F$) or unfairly ($s_G=U$) when interacting with a good recipient. The second letter ($s_B$) represents the dictator's behavior towards a bad recipient, acting fairly ($s_B=F$) or unfairly ($s_B=U$).
Each interaction is observed by a randomly chosen individual, referred to as the observer.
The observer subsequently uses a social norm to establish and spread the reputation of the dictator.
In this study, we utilize a social norm of third-order information, which is represented by a $2\times4$ matrix
\begin{eqnarray*}
\begin{array}{l}
M=\begin{bmatrix}
	f(G,F,G), f(G,U,G), f(B,F,G), f(B,U,G)\\
	f(G,F,B), f(G,U,B), f(B,F,B), f(B,U,B)
\end{bmatrix}
\end{array}.
\end{eqnarray*}
Specifically, $f(x,a,y)=1$ indicates that in the next round, a dictator with the reputation $x$ ($G$ or $B$) will be evaluated as good when taking the action $a$ ($F$ or $U$) towards a recipient with the reputation $y$ ($G$ or $B$). Conversely, $f(x,a,y)=0$ indicates that the dictator will be regarded as bad in the corresponding situation. Since each component in the matrix is binary, we have $2^8=256$ different social norms.

We first study the benchmark (Fig.~\ref{f1}(b)), where it is compulsory for the dictator and the recipient to play the DG. Then the option of voluntary participation is introduced in our model as follows.
In the scenario $1$ (Fig.~\ref{f1}(b)), good dictators are endowed with the option of voluntary participation. Here, good dictators can quit from games with a probability $p_{1}$ when recipients' reputation is bad. If the game does not occur, both the dictator and the recipient receive a small payoff $\sigma$. In the scenario $2$ (Fig.~\ref{f1}(b)), recipients are provided with the option of voluntary participation. Here recipients can quit from games with a probability $p_{2}$ when the dictators are bad. If the game does not occur, both the dictator and the recipient receive no payoffs.
The random role assignment is used to ensure that both players are endowed with an equal probability of being assigned as the dictator or the recipient. If the game does not occur, the reputation of the chosen dictator remains unchanged. Following the common practice, we introduce a small error $\varepsilon$ to account for imperfection in behavior. Particularly, $\varepsilon$ represents the likelihood of a dictator failing to divide the money fairly even when his intention is to do so.

Players use the pairwise comparison rule to update their strategies. In each generation, we randomly choose two players as the model and the imitator, respectively. With a small probability $\mu$, the imitator randomly selects one of the alternative strategies, meaning a mutation event occurs. Otherwise, the imitator $x$ follows the strategy of the model $y$ with the probability
\begin{equation}
	p(x\xrightarrow{}y) = 1/(1+e^{-\beta (g_y-g_x)}).
\end{equation}
Here, $\beta$ represents the selection intensity, $g_x$ and $g_y$ are the payoffs obtained by the imitator and the model, respectively. 
\begin{assumption}\label{assumption_1}
Reputation update occurs significantly faster compared with strategy selection. It means that the reputation system quickly reaches equilibrium before the strategies of the population change.
\end{assumption}

Under Assumption 1, we use two timescales to explore the evolutionary outcomes. As shown in Fig.~1 of Supplementary Information, we first
focus on the reputation update given the strategies of all players remain unchanged. Subsequently, players are allowed to update their strategies over time and we focus on the strategy selection. In the following, we will describe the concrete calculation process of the fairness level, as depicted in Fig.~2 of Supplementary Information.

\subsection{The reputation update}
We assume that all players have fixed strategies and investigate the Markov chain describing the reputation system. Consider a population state with $m$ players of the strategy $X = s^X_Gs^X_B$ and $Z-m$ players of the strategy $Y = s^Y_Gs^Y_B$. Here, there are $(m+1)\times(Z-m+1)$ reputation states in all. A reputation state $(i,j)$ denotes that $i$ of $X-$players are assessed as good and $j$ of $Y-$players are regarded as good. Assume $p(i,j;i^{'},j^{'})$ is the transition probability that the reputation state will change from $(i,j)$ to $(i^{'},j^{'})$ after a round of game. We derive the transition probability matrix $P=(p(i,j;i^{'},j^{'}))$ in Supplementary Information. By calculating the normalized left eigenvector of $P$ corresponding to the eigenvalue $1$,
we get the stationary frequency of each reputation state $(i,j)$. When the population consists of $m$ players of the strategy $X$ and $(Z-m)$ players of the strategy $Y$, the expected payoff obtained by an $X-$player or a $Y-$player is
\begin{equation}
	\begin{split}
		g_X = \sum_{i=0}^{m} \sum_{j=0}^{Z-m} v(i,j) \pi_X(m,i,j), \\
		g_Y = \sum_{i=0}^{m} \sum_{j=0}^{Z-m} v(i,j) \pi_Y(m,i,j).
	\end{split}	
\end{equation}
Here, $\pi_X(m,i,j)$ or $\pi_Y(m,i,j)$ is the expected payoff obtained by an $X-$player or a $Y-$player in the reputation state $(i,j)$. The calculation details for $\pi_X(m,i,j)$ and $\pi_Y(m,i,j)$ are shown in Supplementary Information.

\subsection{The strategy selection and the fairness level}
\begin{assumption}\label{assumption_2}
The mutation rate is extremely low, that is, $\mu \xrightarrow{} 0$. It means that a single mutant quickly fixates or diminishes in the population before the next mutant appears.
\end{assumption}
Under Assumption 2, the evolution of strategies can be approximated by an embedded Markov chain. Since both $s_G$ and $s_B$ are binary, the corresponding state space is comprised of four monomorphic population states. In each monomorphic population state, all players use the same strategy $UU$, $UF$, $FU$, or $FF$. Assume that $\rho _{XY}$ is the probability that the single $X$-mutant fixates in a population with $Z-1$ $Y-$players and a single $X-$player. The detailed calculation of $\rho _{XY}$ is presented in Supplementary Information. Then the transition matrix can be described by $A=(a_{XY})_{4 \times 4}$ where $a_{XY}= \frac{\mu}{4} \rho _{YX}$ with $X \neq Y$ and $a_{XX} = 1-\sum_{Y,Y \neq X}a_{XY}$. The steady-state distribution $\Phi = (\phi_{UU},\phi_{UF},\phi_{FU},\phi_{FF})$ can be derived by calculating the normalized left eigenvector of $A$ corresponding to the eigenvalue $1$, meaning $\Phi = \Phi A$.

The fairness level can be expressed by the steady-state frequencies of strategies as
\begin{equation}
	\label{e1}
	F = \sum_{X \in \{UU,UF,FU,FF\}} \phi _X f_F (X),
\end{equation}
where $f_F(X)$ represents the fairness level observed in the monomorphic population of $X$. Specifically, $f_F(UU) = 0$ since $UU$ dictators always make unfair splits. $f_F(FF) = 1-\varepsilon$ since $FF$ dictators act fairly with a probability $1-\varepsilon$. For $X \in \{FU,UF\}$, $f_F(X)$ is associated with the number of good players, and it can be expressed as
\begin{equation}
	f_F(X) = \sum_{i=0}^{Z} v(i|X)p_F(i|X).
\end{equation}
Here, $v(i|X)$ represents the stationary frequency of the reputation state with $i$ good players in the monomorphic population of $X$. In this case, $v(i|X)$ is equivalent to $v(i,0)$ in the monomorphic population of $X$. $p_F(i|X)$ denotes the probability that a player makes a fair split as the dictator or receives a fair split as the receiver when the monomorphic population of $X$ is comprised of $i$ good players and $Z-i$ bad players. The expression can be given as
\begin{equation}
	\label{e2}
	p_F(i|X) = (1- \varepsilon)[\frac{i}{Z}I(s_G^X)+\frac{Z-i}{Z}I(s_B^X)],
\end{equation}
where $I(F)=1$ and $I(U)=0$.

\section{Results}
\begin{figure*}
	\centering
	\includegraphics[width=\linewidth]{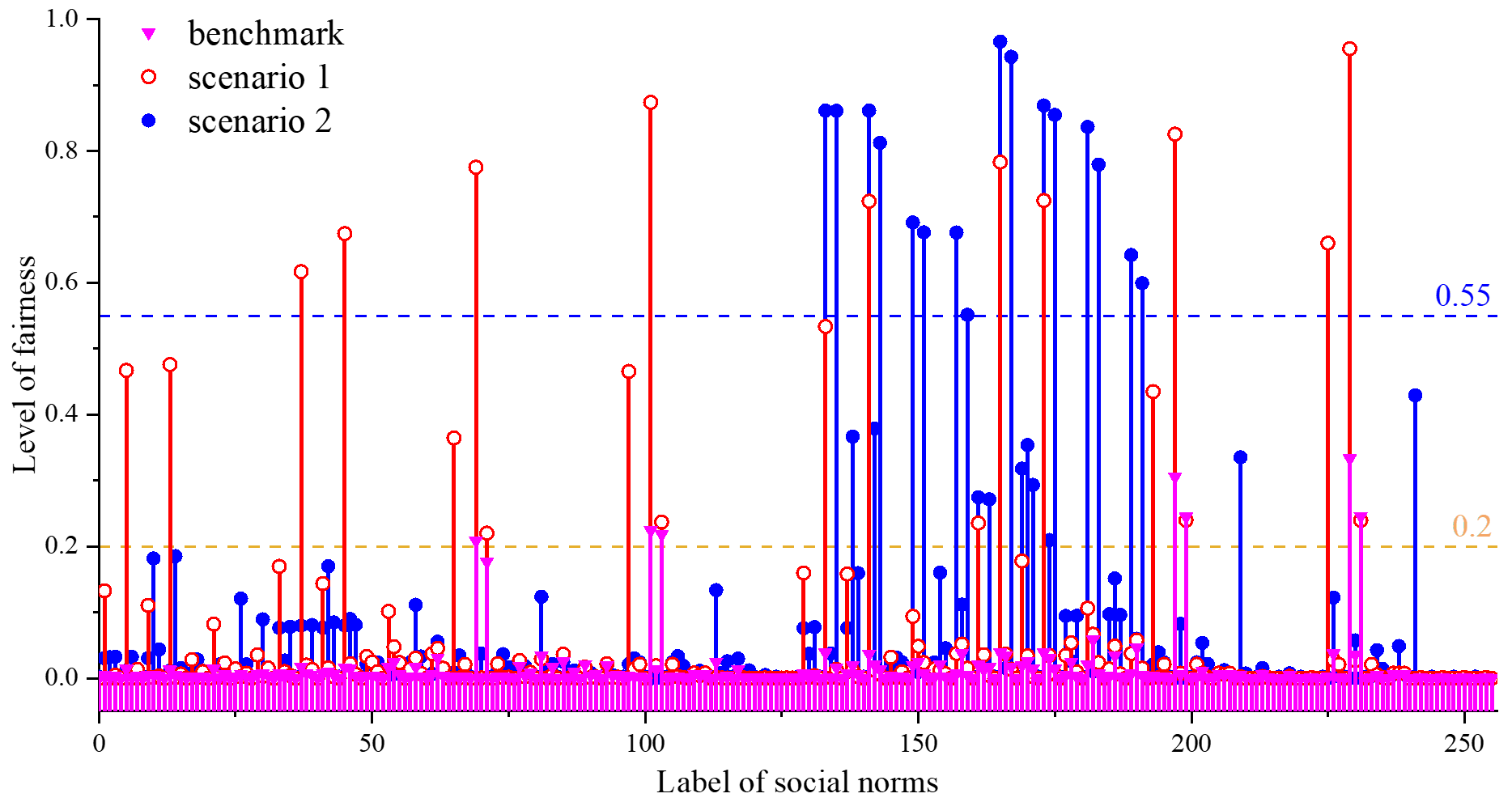}
	\caption{\label{fairness_level} The fairness level under $256$ social norms in different cases. All social norms induce the low fairness level which is less than $34\%$ at the benchmark (pink solid triangle). In the first scenario we considered (red hollow circle), ten social norms lead to the high fairness level greater than $0.55$. In the second scenario we considered (blue solid circle), sixteen social norms result in the high fairness level which is greater than $0.55$. A social norm $M=[f(G,F,G),f(G,U,G),f(B,F,G),f(B,U,G);f(G,F,B),f(G,U,B),f(B,F,B),f(B,U,B)]$ is labelled as the sum $(f(G,F,G)\times2^7+f(G,U,G)\times2^6+f(B,F,G)\times2^5+f(B,U,G)\times2^4+ f(G,F,B)\times 2^3+f(G,U,B)\times 2^2+f(B,F,B)\times 2^1+ f(B,U,B)\times 2^0)$. Parameters: $Z=50$, $\mu=0.01$, $\epsilon=0.01$, $\sigma =0.1$, $p=0.01$, $p_{1}=p_{2}=0.5$, and $\beta=1$.}
\end{figure*}

Based on the above description, we can conduct a comprehensive analysis about how the option of voluntary participation influences the fairness level under $256$ third-order social norms. We are also interested in identifying the social norms that can encourage the emergence of fairness. Then, we elucidate the high fairness level by examining the pairwise competition between all strategies and the reputation update in each monomorphic population. As presented in Fig.~\ref{fairness_level}, all social norms induce the low fairness level which is less than $34\%$ at the benchmark. It implies that reputation alone cannot promote the emergence of fairness behavior.

\subsection{Good dictators with voluntary participation}

\begin{figure*}[t]
	\centering
\includegraphics[width=\linewidth]{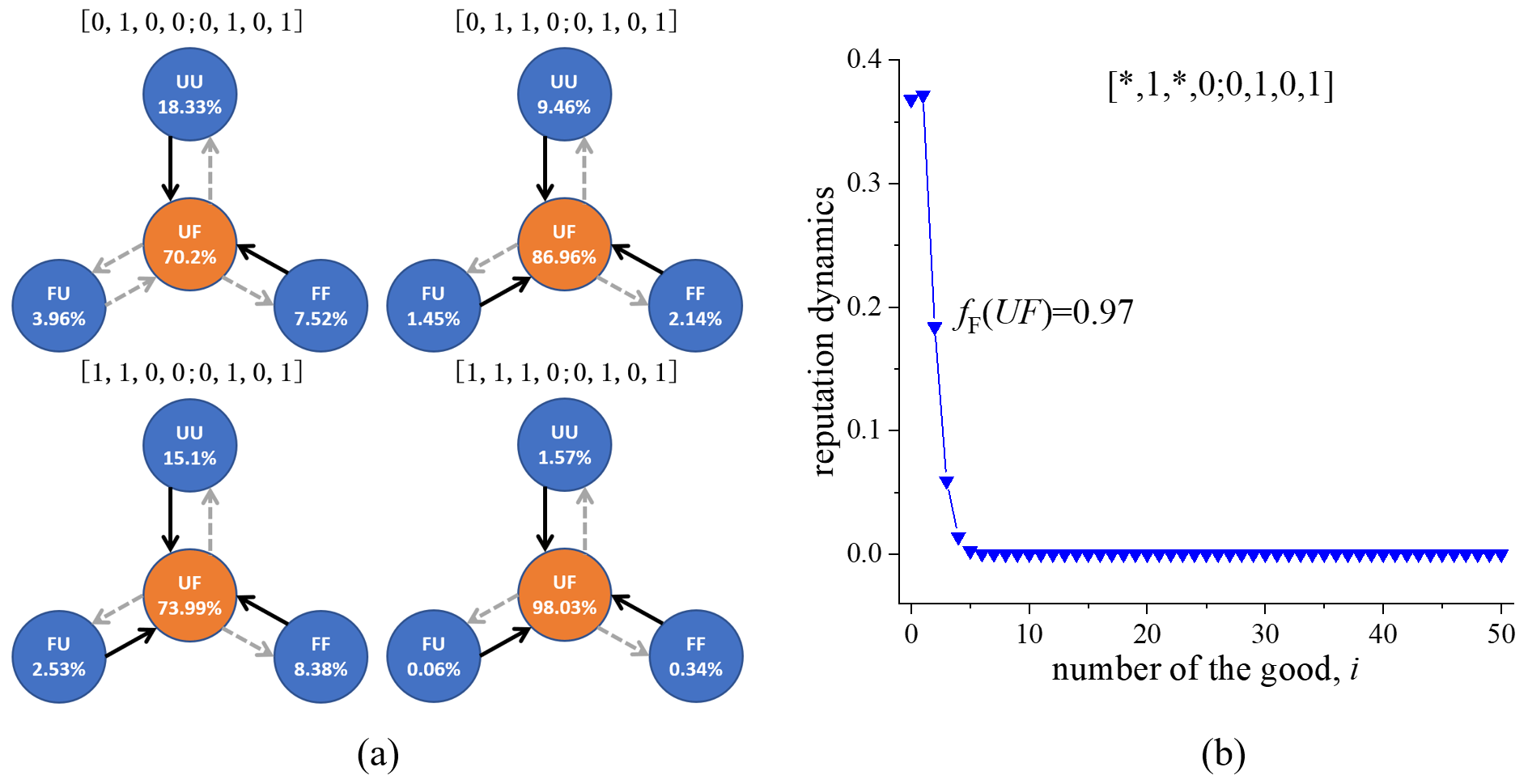}
	\caption{\label{dic_loner} Evolutionary outcomes in the scenario $1$ under the social norms $[*,1,*,0;0,1,0,1]$. (a) The strategies and their frequencies are represented by the letters and the numbers in the circles. When the fixation probability of a single mutant (the ending point) into the given resident strategy (the starting point) is greater than $1/Z$, the arrows are depicted with the solid lines. When the fixation probability is not greater than $1/Z$, the arrows are depicted with the dashed lines. $UF$ has the high frequencies which are greater than $70\%$. It has the superiority over $UU$ and $FF$. Simultaneously it is not favored to be invaded by $FU$. (b) The reputation dynamics for the monomorphic population of $UF$ peaks at the state near `all bad'. Parameters: $Z=50$, $\mu=0.01$, $\epsilon=0.01$, $\sigma =0.1$, $p=0.01$, $p_{1}=0.5$, and $\beta=1$.}
\end{figure*}

\begin{figure*}[t]
	\centering
	\includegraphics[width=\linewidth]{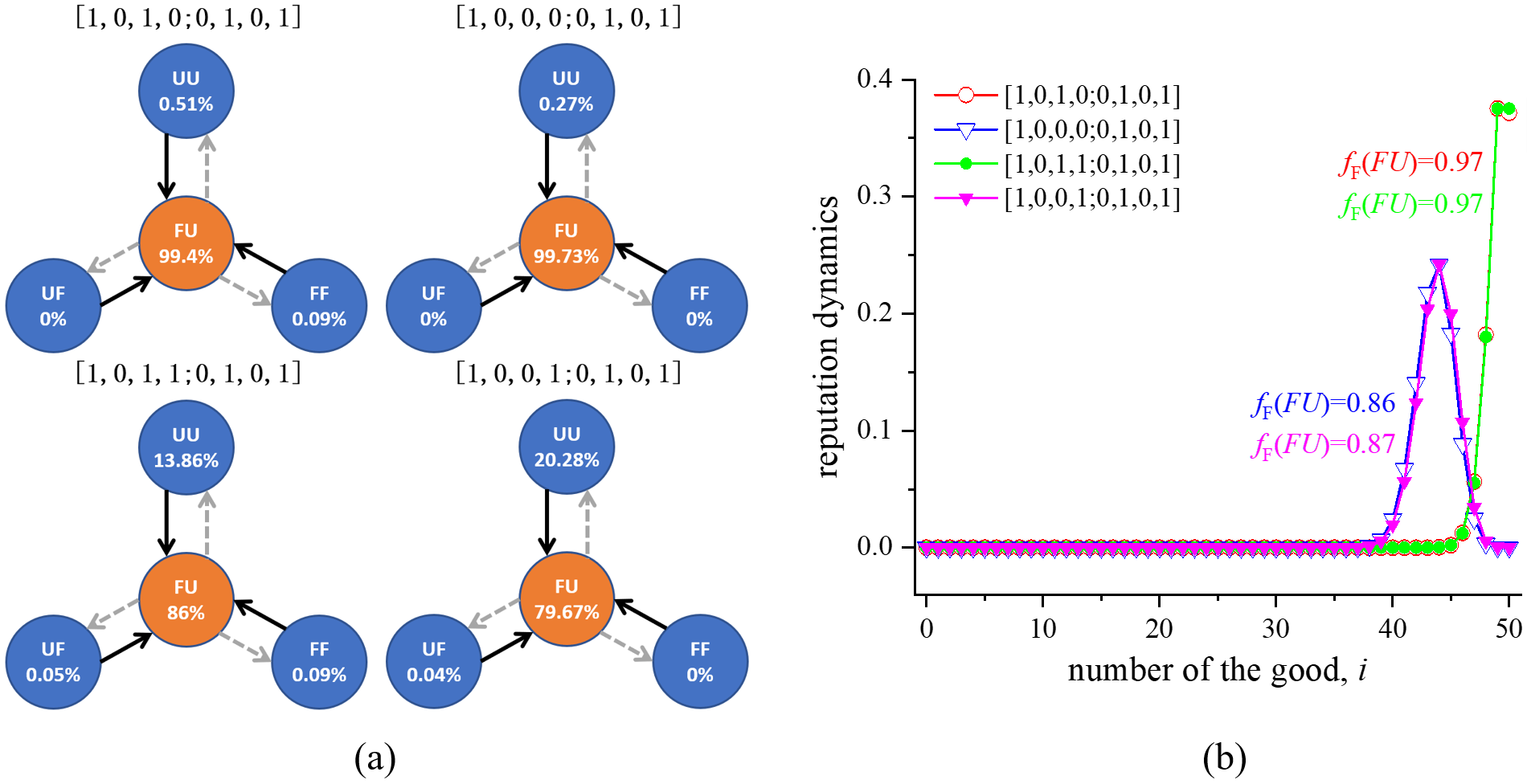}
	\caption{\label{recipient_loner} Evolutionary outcomes in the scenario $2$ under the social norms $[1,0,*,*;0,1,0,1]$. (a) The strategies and their frequencies are represented by the letters and the numbers in the circles. When the fixation probability of a single mutant (the ending point) into the given resident strategy (the starting point) is greater than $1/Z$, the arrows are depicted with the solid lines. When the fixation probability is not greater than $1/Z$, the arrows are depicted with the dashed lines. The steady-state frequencies of $FU$ are consistently above $79\%$. $FU$ is superior to the other three strategies. (b) The reputation dynamics for the monomorphic population of $FU$ peaks at the state near `all good'. Parameters: $Z=50$, $\mu=0.01$, $\epsilon=0.01$, $p=0.01$, $p_{2}=0.5$, and $\beta=1$.}
\end{figure*}
As a punishment imposed on bad players, we endow good dictators with the option of voluntary participation. Here, good dictators can quit from games with a probability $p_{1}$ when interacting with bad recipients. Here, ten social norms which belong to the set $\{[*,*,*,0;*,*,0,1]\}$ ($*$ denotes $0$ or $1$) lead to the high fairness level which is greater than $0.55$ for $p_{1} = 0.5$. Notably, $f(B,U,G)=0$, $f(B,F,B)=0$, and $f(B,U,B)=1$ are necessary but not sufficient conditions to induce the high fairness level greater than $0.55$.

We take the four social norms $[*,1,*,0;0,1,0,1]$ as examples to understand the high fairness level. Eq.~(\ref{e1}) indicates that the fairness level is determined by the stationary frequency of each strategy and the fairness level exhibited in each monomorphic population. The strategy $UF$ has the large frequencies which are greater than $70\%$ (Fig.~\ref{dic_loner}(a)). Meanwhile, the monomorphic population of $UF$ exhibits the high fairness level close to $100\%$ (Fig.~\ref{dic_loner}(b)). Conversely, both $FU$ and $FF$ have quite small frequencies, and they contribute very little to the fairness level in the population. Additionally, the strategy $UU$ does not contribute to the fairness level at all. Therefore, it is evident that the fairness level is primarily driven by $UF$.

When the mutation rate is very low, the stationary frequency of a strategy is determined by its pairwise competition with other strategies. Some definitions are first given to have a clear and concise explanation for this. If the fixation probability of a single $X$ mutant in the population with the resident strategy $Y$ is above (not above) $1/Z$ (neutral fixation probability), then $X$ is (not) favored to invade $Y$. If $X$ is favored to invade $Y$ and $Y$ is not favored to invade $X$, then $X$ is superior to $Y$. In this context, we explain why $UF$ dominates in the population by examining its pairwise competition with the other three strategies. From Fig.~\ref{dic_loner}(a), the large frequencies of $UF$ can be mainly attributed to its superiority over $UU$ and $FF$. Furthermore, $UF$ is not favored to be invaded by $FU$.

The superiority of $UF$ over $UU$ can be understood as follows. In the pairwise competition, we assume that the initial population is comprised of bad players. For $[*,1,*,0;0,1,0,1]$, the $UF$ dictator obtains a bad reputation by the fair division towards bad recipients since $f(\circ,F,B)=0$ ($\circ\in\{G,B\}$). Meanwhile, the $UF$ dictator can maintain the bad reputation by the unfair division towards good recipients since $f(B,U,G)=0$. Yet the $UU$ dictator is assessed as good by the unfair division with bad recipients as indicated by $f(\circ,U,B)=1$.
The $UU$ dictator can sustain the good reputation by the unfair division with good recipients, as indicated by $f(G,U,G)=1$.
When $p_{1}=0.5$, $UU$ dictators with good reputation have the option of quitting from games with $UF$ recipients who are typically perceived as bad. Then $UU$ dictators receive a very small payoff $\sigma=0.1$. However, $UF$ dictators with bad reputation cannot quit from games, and get a minimal profit of $0.5$. Furthermore, $UF$ recipients with bad reputation can expect fair splits from $UF$ dictators. Yet $UU$ recipients with good reputation consistently receive unfair splits regardless of their opponents. Consequently, the strategy $UF$ possesses a payoff advantage over the strategy $UU$.
Similarly, we can analyze why $UF$ is superior to $FF$. The $FF$ dictator can easily acquire and maintain a bad reputation by the fair division towards bad recipients ($f(\circ,F,B)=0$). $FF$ players with bad reputation have the same payoff as $UF$ players with bad reputation. However, there still exist a few good players in the population due to the mistake implementation ($f(\circ,U,B)=1)$. As a result, $UF$ dictators keep almost all of the sum to themselves when interacting with good recipients. On the other hand, $FF$ dictators consistently receive a payoff of $0.5$ regardless of their opponents' reputation. Therefore, $UF$ maintains a payoff advantage over $FF$.

We can understand why $UF$ cannot be invaded by $FU$ as follows. The above analysis shows that $UF$ players are typically regarded as bad for $[*,1,*,0;0,1,0,1]$. When the number of $FU$ players is small, the $FU$ dictator can easily attain a good reputation by the unfair division with bad recipients ($f(\circ,U,B)=1$). Here, the $FU$ dictator with a good reputation has the option of voluntary participation when facing a bad $UF$ recipient. Then the $FU$ dictator obtains a small payoff $\sigma=0.1$. On the other hand, the $UF$ dictator with a bad reputation is unable to quit from games and therefore acts unfairly against the good $FU$ recipient. Then the $UF$ dictator keeps almost all of the sum to himself. Consequently, $UF$ has a payoff advantage over $FU$.

We also explore why the high fairness level is observed in the monomorphic population of $UF$. Fig.~\ref{dic_loner}(b) shows that the four social norms $[*,1,*,0;0,1,0,1]$ exhibit an identical reputation dynamics. This is because they share the same values of $f(\circ,U,G)$ and $f(\circ,\bullet,B)$ ($\bullet\in\{F,U\}$). Besides, $UF$ means acting unfairly against good recipients and acting fairly or unfairly against bad recipients. For $[*,1,*,0;0,1,0,1]$, the $UF$ dictator tends to obtain bad reputation by acting fairly against bad recipients ($f(\circ,F,B)=0$). Meanwhile, the $UF$ dictator can keep the bad reputation by acting unfairly against good recipients ($f(B,U,G)=0$). Then the stationary distribution of the reputation system has a single peak at the state near `all bad' (Fig.~\ref{dic_loner}(b)). Furthermore, the monomorphic population of $UF$ exhibits the high fairness level.

\subsection{Recipients with voluntary participation}

When good dictators are endowed with voluntary participation, the high fairness level is induced by acting fairly against bad recipients. Obviously, it is not worth being encouraged in the real life. Then we explore an alternative punishment mechanism by endowing recipients with voluntary participation. Here, recipients can quit from games with a probability $p_{2}$ when interacting with bad dictators. If the game does not occur, neither the dictator nor the recipient receives any payoffs. For $p_{2} = 0.5$, sixteen social norms of the set $\{[1,0,*,*;*,1,*,1]\}$ correspond to the high fairness level greater than $0.55$ (Fig.~\ref{fairness_level}). Specifically, $f(G,F,G)=1$, $f(G,U,G)=0$, and $f(*,U,B)=1$ are sufficient and necessary conditions for the high fairness level greater than $0.55$.

The four social norms $[1,0,*,*;0,1,0,1]$ are taken as examples to understand the high fairness level in the population. The steady-state frequencies of $FU$ are consistently above $79\%$ (Fig.~\ref{recipient_loner}(a)). The monomorphic population of $FU$ exhibits a high fairness level (Fig.~\ref{recipient_loner}(b)). Both $UF$ and $FF$ have extremely low steady-state frequencies, and they have little effect on the fairness level in the population. Additionally, $UU$ does not contribute at all to the fairness level. Consequently, the fairness behavior primarily stems from $FU$. The reason why $FU$ dominates in the population is that $FU$ is superior to the other three strategies (Fig.~\ref{recipient_loner}(a)).

The superiority of $FU$ over $UU$, $UF$, or $FF$ can be understood as follows. In the pairwise competition, the initial population is comprised of bad players. For $[1,0,*,*,0,1,0,1]$, the $FU$ dictator is regarded as good by the unfair division towards bad recipients ($f(\circ,U,B)=1$). Meanwhile, the $FU$ dictator can keep the good reputation by the fair division towards good recipients ($f(G,F,G)=1$). 
Yet the $UU$ or $UF$ dictator is likely to be evaluated as bad by the unfair division towards good recipients ($f(G,U,G)=0$). The $UF$ or $FF$ dictator can easily acquire a bad reputation by the fair action towards bad recipients ($f(\circ,F,B)=0$). $FU$ dictators with good reputation can guarantee a minimal profit of $0.5$. Conversely, $UU$, $UF$ or $FF$ dictators with bad reputation may not receive any payoffs because recipients can quit from games with bad dictators. Furthermore, $FU$ recipients with good reputation can expect fair splits from the $FU$ dictators. Yet $UU$, $UF$ or $FF$ recipients with bad reputation receive unfair splits from the $FU$ dictators. Accordingly, $FU$ possesses the payoff advantage over $UU$, $UF$, or $FF$.

We also conduct an investigation of the reasons why the high fairness level is observed in the monomorphic population of $FU$. For $[1,0,1,*;0,1,0,1]$, $FU$ dictators are regarded as good by the unfair division with bad recipients ($f(\circ,U,B)=1$).
Meanwhile, $FU$ dictators obtain good reputation by the fair division with good recipients ($f(\circ,F,G)=1$). Then the state of `all good' is the peak of the reputation system for the monomorphic population of $FU$ (Fig.~\ref{recipient_loner}(b)). By providing fair offers to good recipients, the corresponding fairness level reaches approximately $ 97\% $. However for $[1,0,0,*;0,1,0,1]$, the bad $FU$ dictators are unable to obtain good reputation by the fair division towards good recipients ($f(B,F,G)=0$). As a consequence, good players become less in the monomorphic population of $FU$. The stationary distribution of the reputation system has a peak at the state of `most good' (Fig.~\ref{recipient_loner}(b)). Then the corresponding fairness level declines to around $86\%$.

\subsection{Probability of voluntary participation}
\begin{figure*}
	\centering
	\includegraphics[width=\linewidth]{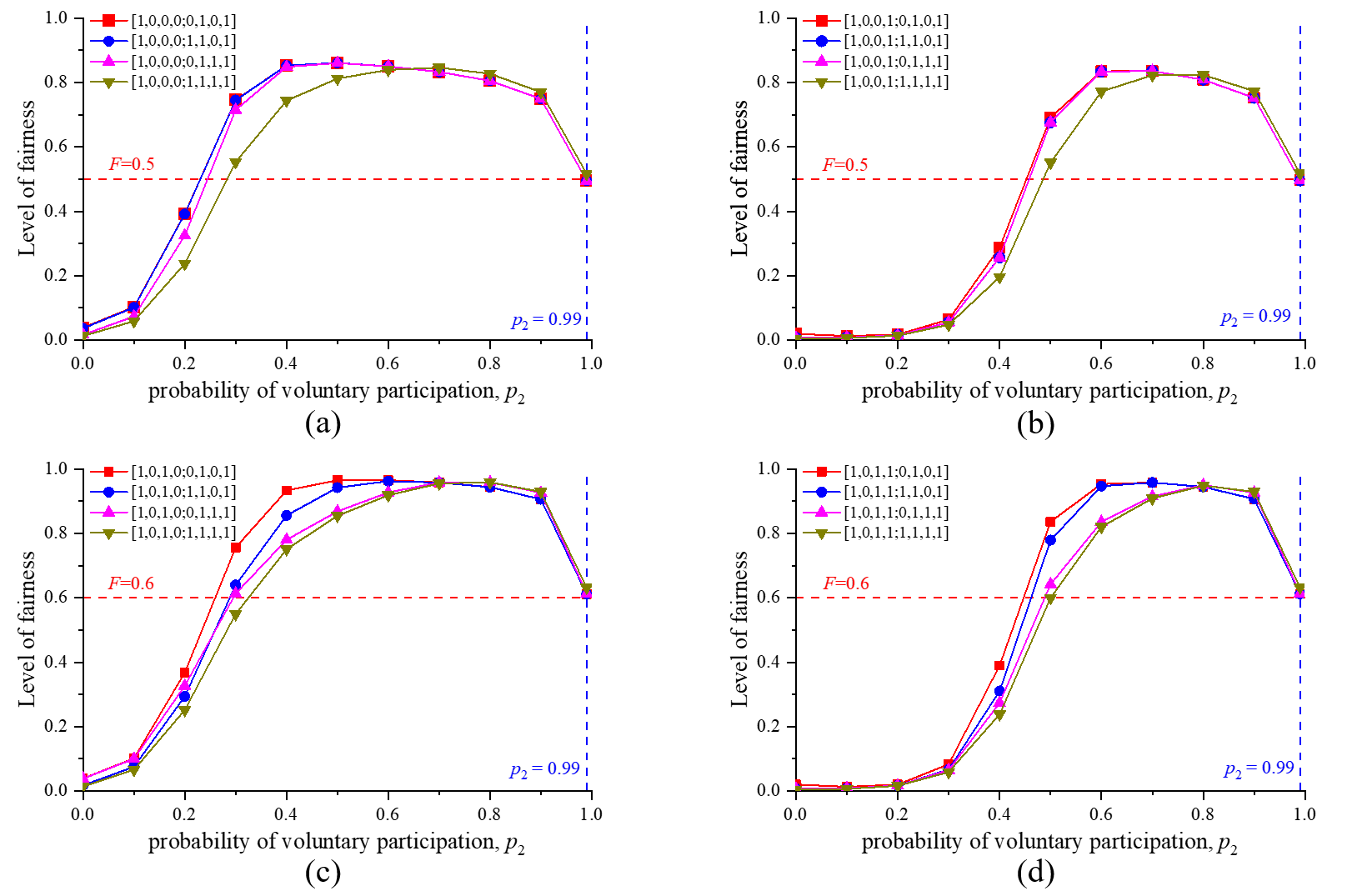}
	\caption{\label{frequency_stra} Under the social norms $[1,0,*,*;*,1,*,1]$, the fairness level initially increases and then decreases with respect to the probability of voluntary participation $p_{2}$. (a) Under $[1,0,0,0;*1,*,1]$, the changing trend of the fairness level with respect to $p_2$. (b) Under $[1,0,0,1;*1,*,1]$, the changing trend of the fairness level with respect to $p_2$. (c) Under $[1,0,1,0;*1,*,1]$, the changing trend of the fairness level with respect to $p_2$. (d) Under $[1,0,1,1;*1,*,1]$, the changing trend of the fairness level with respect to $p_2$. Parameters: $Z=50$, $\mu=0.01$, $\epsilon=0.01$, $p=0.01$, and $\beta=1$.}
\end{figure*}
We present the change of the fairness level with respect to $p_{2}$ under the social norms $[1,0,*,*;*,1,*,1]$ in Fig.~\ref{frequency_stra}. As $p_{2}$ increases, the fairness level initially rises and then declines. When $p_2$ goes away from $0$, the initial growth of the fairness level is fast for $[1,0,*,0;*,1,*,1]$
(Figs.~\ref{frequency_stra}(a) and~\ref{frequency_stra}(c)) and is slow for $[1,0,*,1;*,1,*,1]$ (Figs.~\ref{frequency_stra}(b) and~\ref{frequency_stra}(d)). The key factor affecting the growth rate in the fairness level is the value of $f(B,U,G)$.
For $[1,0,0,*;*,1,*,1]$, the maximal value does not exceed $0.9$ and the fairness level gets close to $0.5$ for $p_{2}=0.99$ (Figs.~\ref{frequency_stra}(a) and~\ref{frequency_stra}(b)). Yet for $[1,0,1,*;*,1,*,1]$, the maximal value is above $0.9$ and the fairness level comes close to $0.6$ for $p_{2}=0.99$ (Figs.~\ref{frequency_stra}(c) and~\ref{frequency_stra}(d)).

\begin{figure*}[h]
	\centering
	\includegraphics[width=\linewidth]{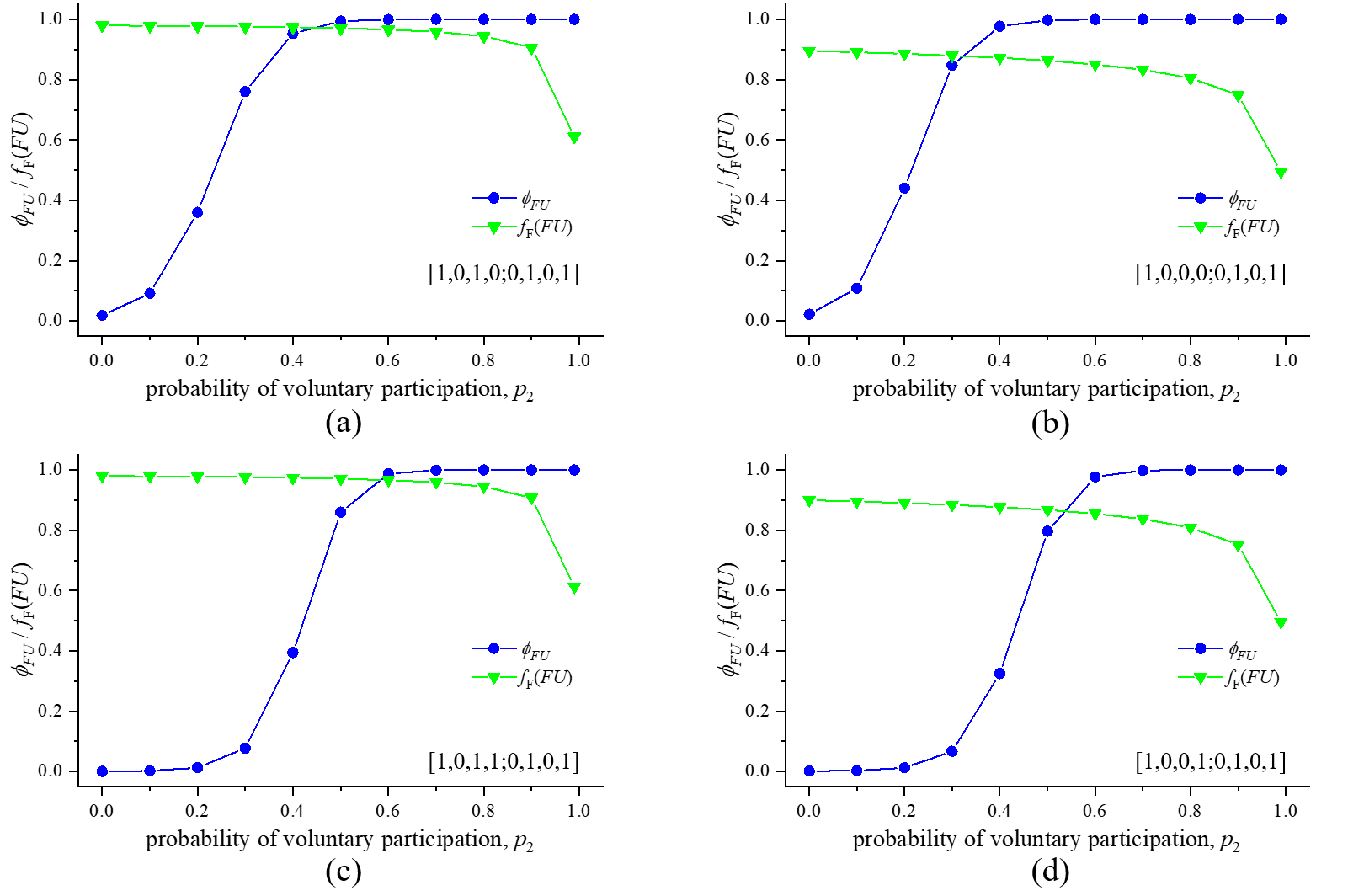}
	\caption{\label{rec_rep} Under the social norms $[1,0,*,*;0,1,0,1]$, the frequency of $FU$ ($\phi_{FU}$) first increases and then changes little with respect to the probability of voluntary participation $p_2$. The fairness level exhibited in the corresponding monomorphic population ($f_F(FU)$) first changes little and then decreases with respect to $p_2$.
	(a) Under $[1,0,1,0;0,1,0,1]$ the changing trend of $\phi_{FU}$ or $f_F(FU)$ with respect to $p_2$. (b) Under $[1,0,0,0;0,1,0,1]$ the changing trend of $\phi_{FU}$ or $f_F(FU)$ with respect to $p_2$.
	(c) Under $[1,0,1,1;0,1,0,1]$ the changing trend of $\phi_{FU}$ or $f_F(FU)$ with respect to $p_2$. (d) Under $[1,0,0,1;0,1,0,1]$ the changing trend of $\phi_{FU}$ or $f_F(FU)$ with respect to $p_2$.
	Parameters: $Z=50$, $\mu=0.01$, $\epsilon=0.01$, $p=0.01$, $p_{2}=0.5$, and $\beta=1$.}
\end{figure*}

We take $[1,0,*,*;0,1,0,1]$ as examples to show the reasons why the fairness level first increases and then decreases with $p_{2}$  (Fig.~\ref{rec_rep}). As $p_{2}$ grows from $0$ to $0.6$, the steady-state frequency of $FU$ quickly increases and the fairness level observed in the monomorphic population changes little. Then the fairness level quickly increases for $p_{2}\in(0, 0.6)$. Yet as $p_2$ grows away from $0.6$, the steady-state frequency of $FU$ remains nearly unchanged and the fairness level observed in the monomorphic population significantly decreases. Consequently, the fairness level undergoes a substantial decline for $p_{2}\in(0.6, 0.99)$.

The reason why the steady-state frequency of $FU$ increases with $p_{2}$ is as follows. As $p_{2}$ initially increases, the punishment inflicted on bad players becomes more severe. Then $FU$ has a greater payoff advantage over the other three strategies.
The underlying cause for the decreasing fairness level in the monomorphic population with $p_2$ is as follows. As $p_{2}$ further increases, recipients become more inclined to abstain from the game with bad dictators. Then, it is more challenging for bad dictators to attain good reputation. Since $FU$ makes an unfair split towards bad players, the increase of bad players results in the deterioration of the fairness level.

In all the analyses of this paper, the value of the selection intensity $\delta$ is fixed to 1. We also investigate other selection intensities in Fig.~3 of Supplementary Information. We find that the social norms which generate the high fairness level for $\delta$ also correspond to the high fairness level for other $\delta$. It implies our main result is robust to the selection intensity.

\section{Conclusions}
We use the indirect reciprocity theory to explore the underlying causes for the emergence of fairness in the evolutionary DG.
Our main findings can be summarized as follows.
	\begin{itemize}
		\item{When both the dictator and the recipient are compelled to participate in the game, all social norms induce the low fairness level less than $34\%$. The result is consistent with previous research~\cite{Li2022}. }
		\item{When good dictators have the option of abstaining from games with bad recipients, ten social norms generate a high fairness level greater than $0.55$ among the $256$ social norms. The fairness behavior primarily stems from the $UF$ strategy. Yet $UF$ exhibits the fairness behavior towards bad recipients. }
		\item {When recipients are given the option of quitting from games with bad dictators, sixteen social norms lead to the high fairness level exceeding $0.55$. Moreover, $f(G,F,G)=1$, $f(G,U,G)=0$, and $f(*,U,B)=1$ are sufficient and necessary conditions for such high fairness level. Here, the fairness behavior is mainly attributed to the $FU$ strategy. Here, dictators perform fairness behavior towards good recipients.}
		\item{For voluntary participation of recipients, $FU$ dictators are typically assessed as good. Yet dictators of the other three strategies are likely to be regarded as bad. Then $FU$ dominates in the population.}
		\item{As the probability of recipients' voluntary participation grows, the fairness level first increases and then decreases.
		The option of voluntary participation can be seen as a form of altruistic punishment~\cite{Sekiguchi2008,Sun2022}. The initial increase of the fairness level is because the punishment for $UU$, $UF$ and $FF$ becomes more severe. The later decrease of the fairness level is because bad dictators are difficult to attain good reputation in the monomorphic population of $FU$.} 
\end{itemize}

Notably, several mechanisms have been proposed to enhance fairness in the DG, including degree-based role assignment~\cite{Deng2022} and multilevel selection~\cite{Jeffrey2015}. However, these studies have been primarily conducted within specific structured populations. Similar to Ref.~\cite{Li2022}, our study focuses on the DG in a well-mixed population to eliminate the influence of network reciprocity. However, in Ref.~\cite{Li2022}, they showed that reputation-based role assignment corresponds to a high fairness level under four of the `leading eight' norms. They investigated the `leading eight' norms and all second-order social norms by a method with two timescales. Differently, we use the same theoretical method to explore how volunteer participation impacts the emergence of fairness under all $256$ third-order social norms. We believe that our results present more comprehensive insights into how fairness emerges in the well-mixed populations without additional interaction structures. Zisis et al. studied the DG where receivers have the option of reputation-based voluntary participation by using the evolutionary game method and behavioral experiments~\cite{Zisis2015}. Besides voluntary participation of receivers, we also consider the scenario where good dictators are endowed with reputation-based voluntary participation. Moreover, we focus on which social norms can promote the evolution of fairness.

Compared with the DG, more studies have been conducted on how fairness evolves within the context of the ultimatum game (UG)~\cite{Zhang2023,Zhao2020,Zheng2022}. In the UG, the recipient possesses the ability to reject low offers and then both participants receive nothing. Yet in the DG, the recipient lacks the ability to refuse unfair splits. Consequently, the fairness level observed in the UG is generally higher than that in the DG. This is consistent with the findings in the UG~\cite{Zhang2023} and the DG of our study. In the UG~\cite{Zhang2023}, four social norms can lead to the fairness level greater than $80\%$ when the proposer updates reputation. In the DG of our study, all social norms correspond to the fairness level less than $34\%$ when the dictator updates reputation.

In this paper, voluntary participation is performed based on reputation. The establishment of a reputation system often entails a cost for observers spreading the reputation of others~\cite{Ohtsuki2015,Sasaki2016,Santos2018}. 
Besides reputation, voluntary participation can also be performed through prior commitment formation~\cite{Han2013,Han2022}.
Arranging a commitment usually also leads to a cost.
Therefore, the impact of the cost of enabling the option of voluntary participation is worthy of investigation in the future.
However, in this work we assume that observers can spread reputation without incurring any costs. Indeed, our theoretical method can be easily extended to the case where observers have to pay a cost to spread reputation. By incorporating this aspect, we predict that more comprehensive insights can be gained for understanding the evolution of fairness behavior in the realistic scenario.

Our findings can provide behavioral scientists or policymakers with norm-based interventions to foster fairness behavior in real life~\cite{Thai2016}. Particularly, in artificial intelligence, the norm-based interventions we proposed can constitute an appealing tool to efficiently steer fairness behavior towards desirable states.

\begin{IEEEbiography}[{\includegraphics[width=1in,height=1.25in,clip,keepaspectratio]{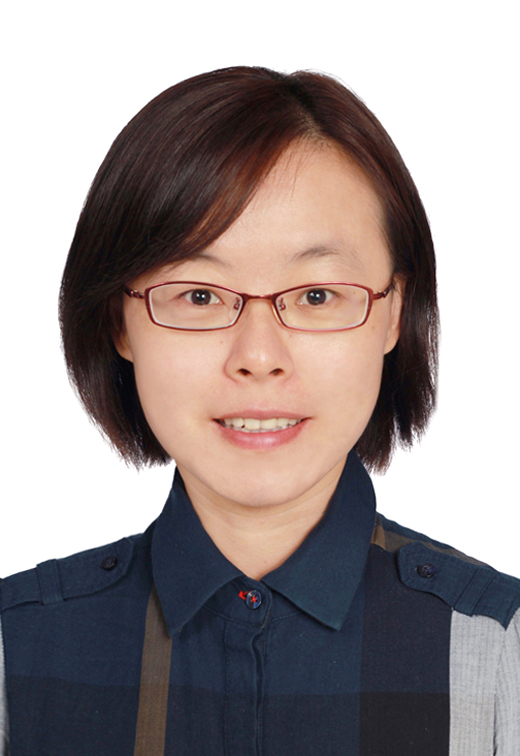}}]{Yanling Zhang}received the B.S. degree in 2009 from Xidian University, China, and the Ph.D. degree in 2015 from Peking University, China. From February 2016 to October 2016, she was a visiting scholar in Dartmouth College, USA.\\
She is currently an Associate Professor at the School of Intelligence Science and Technology in University of Science and Technology Beijing, China. Her main research interests include evolutionary game theory, collective intelligence, and game learning. She has published over 30 journal papers.\end{IEEEbiography}

\begin{IEEEbiography}[{\includegraphics[width=1in,height=1.25in,clip,keepaspectratio]{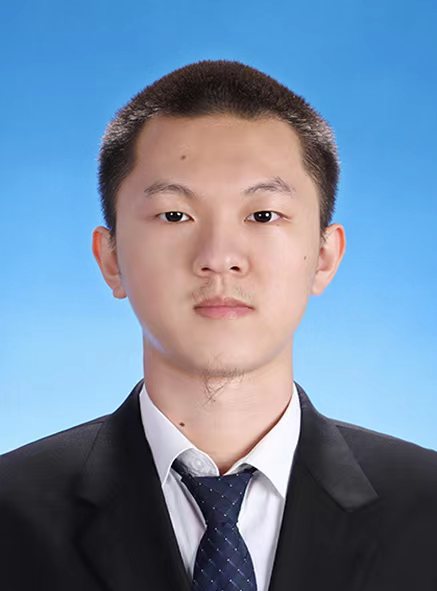}}]{Yin Li}received the B.S. degree in 2022 from the School of Automation and Electrical Engineering, University of Science and Technology Beijing, China. \\
He is currently pursuing the M.S. degree from the School of Intelligence Science and Technology, University of Science and Technology Beijing, China. His current research interests include evolutionary game theory and game learning.
\end{IEEEbiography}

\begin{IEEEbiography}[{\includegraphics[width=1in,height=1.25in,clip,keepaspectratio]{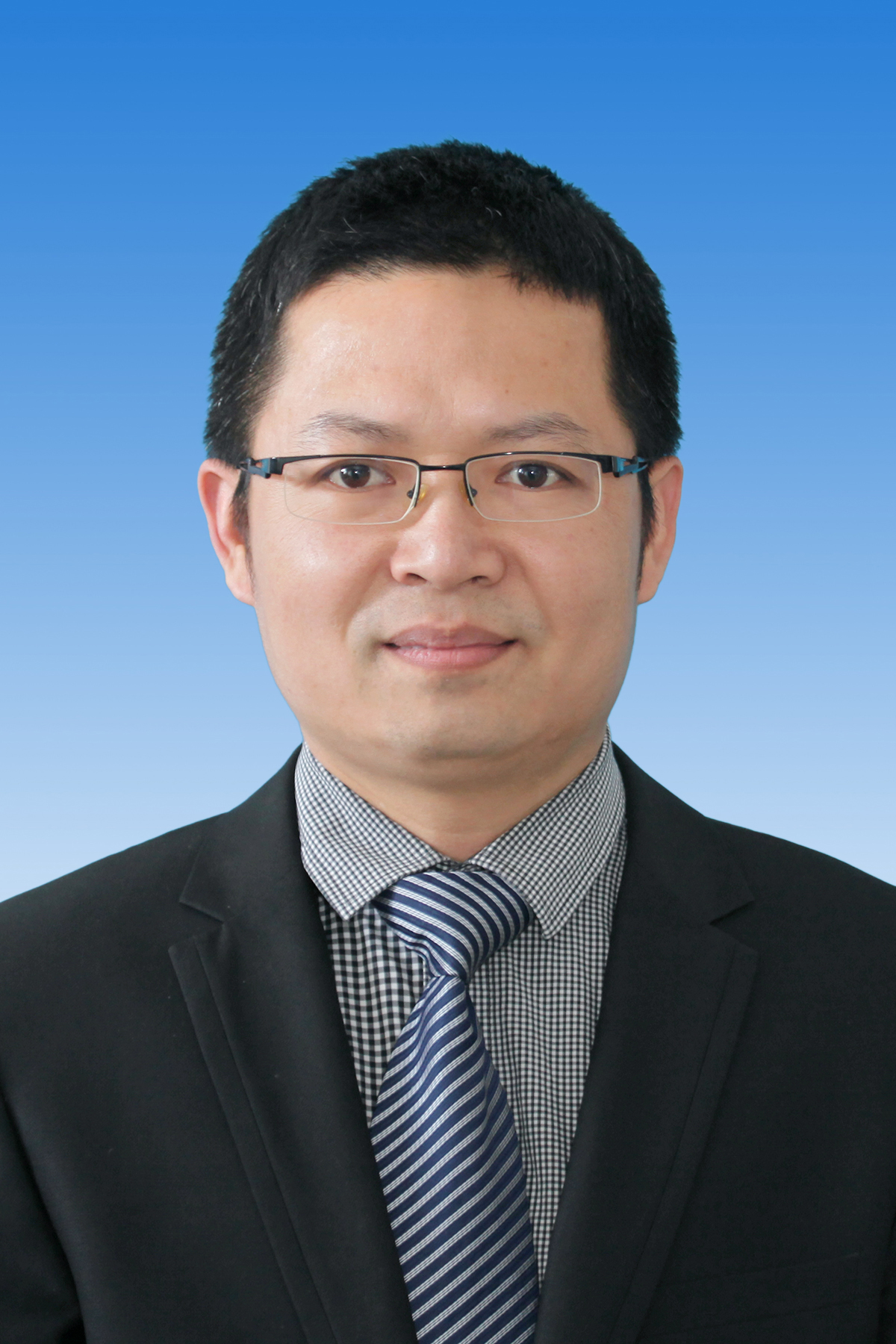}}]{Xiaojie Chen}received the B.S. degree in 2005 from National University of Defense Technology, China, and the PhD degree in 2011 from Peking University, China. From September 2008 to September 2009, he was a visiting scholar in University of British Columbia, Canada. From February 2011 to January 2013, he was a postdoctoral research scholar at the International Institute for Applied Systems Analysis (IIASA), Austria. From February 2013 to January 2014, he was a research scholar at IIASA, Austria.\\
He is currently a professor at the School of Mathematical Sciences in University of Electronic Science and Technology of China, China. His main research interests include evolutionary game dynamics, decision-making in game interactions, game-theoretical control, and collective intelligence. He has published over $100$ journal papers.
\end{IEEEbiography}

\begin{IEEEbiography}[{\includegraphics[width=1in,height=1.25in,clip,keepaspectratio]{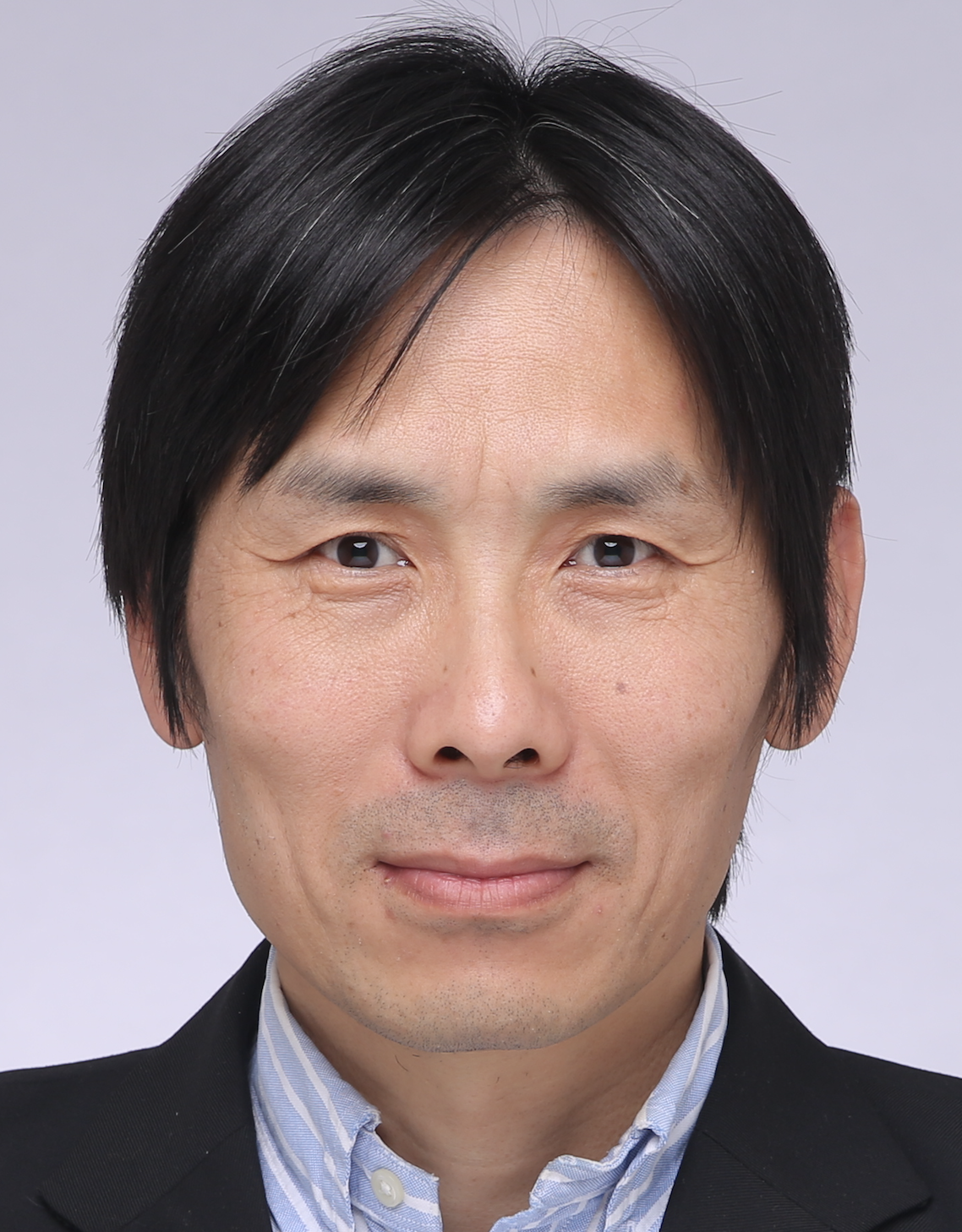}}]{Guangming Xie}received the B.S. degrees in both Applied Mathematics and Electronic and Computer Technology, the M. E. degree in Control Theory and Control Engineering, and the Ph.D. degree in Control Theory and Control Engineering from Tsinghua University, China in 1996, 1998, and 2001, respectively. Then he worked as a postdoctoral research fellow in the Center for Systems and Control, Peking University, China from July 2001 to June 2003.\\ 
He is currently a full Professor of the College of Engineering, Peking University and the founder and director of the Intelligent Biomimetic Design Lab (IBDL). He is the main editor of Mathematical Problems in Engineering and serves as associate editors of several international journals. His research interests include biomimetic robot and multi-robot cooperation. He has published over 200 papers and three books. He is one of the Most Cited Chinese Researchers from Elsevier (Control and system engineering), from 2014 to 2022.
\end{IEEEbiography}

\vfill

\end{document}